\documentstyle[12pt,epsf,here,a4wide,fancyheadings]{article}
\def \be {\begin{equation}}
\def \e {\end{equation}}
\def \bea {\begin{eqnarray}}
\def \ea {\end{eqnarray}}
\def \ep {\epsilon}

\def \no {\nonumber}

\def \sub {\scriptscriptstyle}

\newcommand{\To}[2]{\stackrel{#1}{\hbox to #2 pt{\rightarrowfill}}}

\def \vector#1{\stackrel{\hspace{-0.45em}\longrightarrow}{#1}}

\def\np#1#2#3{{\it Nucl.~Phys.\/}~{\bf B#1} (19#2) #3}
\def\pl#1#2#3{{\it Phys.~Lett.\/}~{\bf B#1} (19#2) #3}

\def\prd#1#2#3{{\it Phys.~Rev.\/}~{\bf D#1} (19#2) #3}
\def\prl#1#2#3{{\it Phys.~Rev.~Lett.\/}~{\bf #1} (19#2) #3}

\def\epj#1#2#3{{\it Eur.~Phys.~J.\/}~{\bf C#1} (19#2) #3}
\begin{document}

\vspace*{1cm}
\begin{center}
{\Large {\bf {Radiation Zeros as an observable to test physics beyond the standard model}}}\\
\vspace{1cm}
{\bf Anja Werthenbach } \\
\vspace{0.6cm}
Department of Physics, University of Durham \\
South Road, Durham DH1 3LE, UK\\

\end{center}
\vspace{2.3cm}

\begin{center}

{\large {\bf {Contents:}}} \\
\end{center}

\vspace{0.5cm}

\hspace{3cm}
\begin{minipage}[r]{15cm}

{ 1. Introduction}\\

{ 2. A bit of History}\\

{ 3. Double W Production}\\

{ 4. Anomalous gauge boson couplings}\\

{ 5. Conclusion}\\
\end{minipage}

\section{Introduction}
Radiation Zeros (RAZ) have first been found in the late seventies by K.O. Mikaelian, M.A. Samuel and D. Sahdev 
\cite{first}. RAZ are points in phase space where for a given process in which a massless gauge boson is emitted the cross section vanishes identically. The origin of RAZ is a cancellation (destructive interference of radiation patterns) which is related to the charge of
the participating particles and only occurs for couplings given within the standard model \cite{brown}. Therefore they are extremly sensitive to physics beyond the standard model.\\  

Radiation Zeros are the generalization of the vanishing of classical nonrelativistic electric and magnetic
dipole radiation occuring for equal mass to charge ratios.\\

\section{A bit of history}

The process $ q \bar{q} \, {'} \to W^+ \gamma$ where a zero was first found can be represented through three
Feynman diagrams, which is from left to right an s-channel, t-channel and u-channel exchange.\\

\begin{figure}[H]
\vskip -2.8cm
\hspace{1.3cm}\centerline{\epsfysize=6cm\epsffile{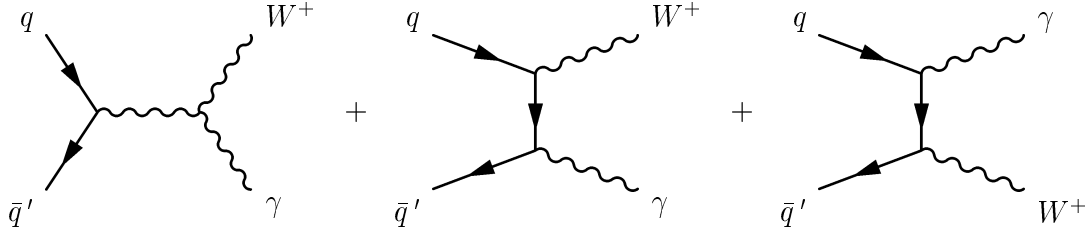}}
\caption{\label{single} {Contributing diagrams to the process $ q \bar{q}{\, '} \to W^+ \gamma$.}}
\end{figure}

Since the cross section $\sigma \sim |{\cal M}|^2$ radiation zeros will occur whenever 

\be 
{\cal M} =  {\cal M}_s + {\cal M}_t + {\cal M}_u = 0  
\label{raz}
\e 

which is realized for the scattering angle of the photon being

\bea
\label{raz}
\fbox{\parbox{9.5cm} {
\bea \cos \theta _{\gamma} \,^{\scriptscriptstyle {RAZ}} = 1-2Q_q = -\frac{1}{3}  \no \ea }}
\ea

with  $Q_q=2/3$, the charge of the $u$ quark.\\

The kinematics used is

\bea
p_1^{\phantom{1}\mu} \, &=& \, ( E, 0, 0, E) \no \\
p_2^{\phantom{2}\mu} \, &=& \, ( E, 0, 0, -E) \no \\
k_+^{\phantom{+}\mu} \, &=& \, \left( \frac{4\, E^2+M_{{\sub W}}^2}{4\, E}, \frac{4\, E^2-M_{\sub W}^2}{4\, E}\,
 \sin \, \Theta \, , \, 0 \, , \, \frac{4\, E^2+M_{\sub W}^2}{4\, E} \cos \, \Theta \right) \no \\
k^{\mu} \phantom{+} \, &=& \, \left( \frac{4\, E^2-M_{\sub W}^2}{4\, E}, - \frac{4\, E^2-M_{\sub W}^2}{4\, E}\,
 \sin \, \Theta \, , \, 0 \, , \, -\frac{4\, E^2+M_{\sub W}^2}{4\, E} \cos \, \Theta \right) \no \\
\ea

with $ \Theta $ being the angle between the incoming quark and the $W^+$ and therefore $\theta_{\gamma} = \Theta +\pi $, 
E being the beam energy of the scattering particles.\\

In 1995 the CDF collaboration \cite{cdf} managed to verify the above result experimentally.\\

\section{Double $W$ production}

The process we are actually interested in is $ q \bar{q} \to W^+W^-\gamma \to f_1 \bar{f_2} f_3 \bar{f_4} \gamma$. 
The contributing feynman diagrams are the following:

\begin{figure}[H]
\vskip -1.8cm
\label{all}
\centerline{ \epsfysize=20cm\epsffile{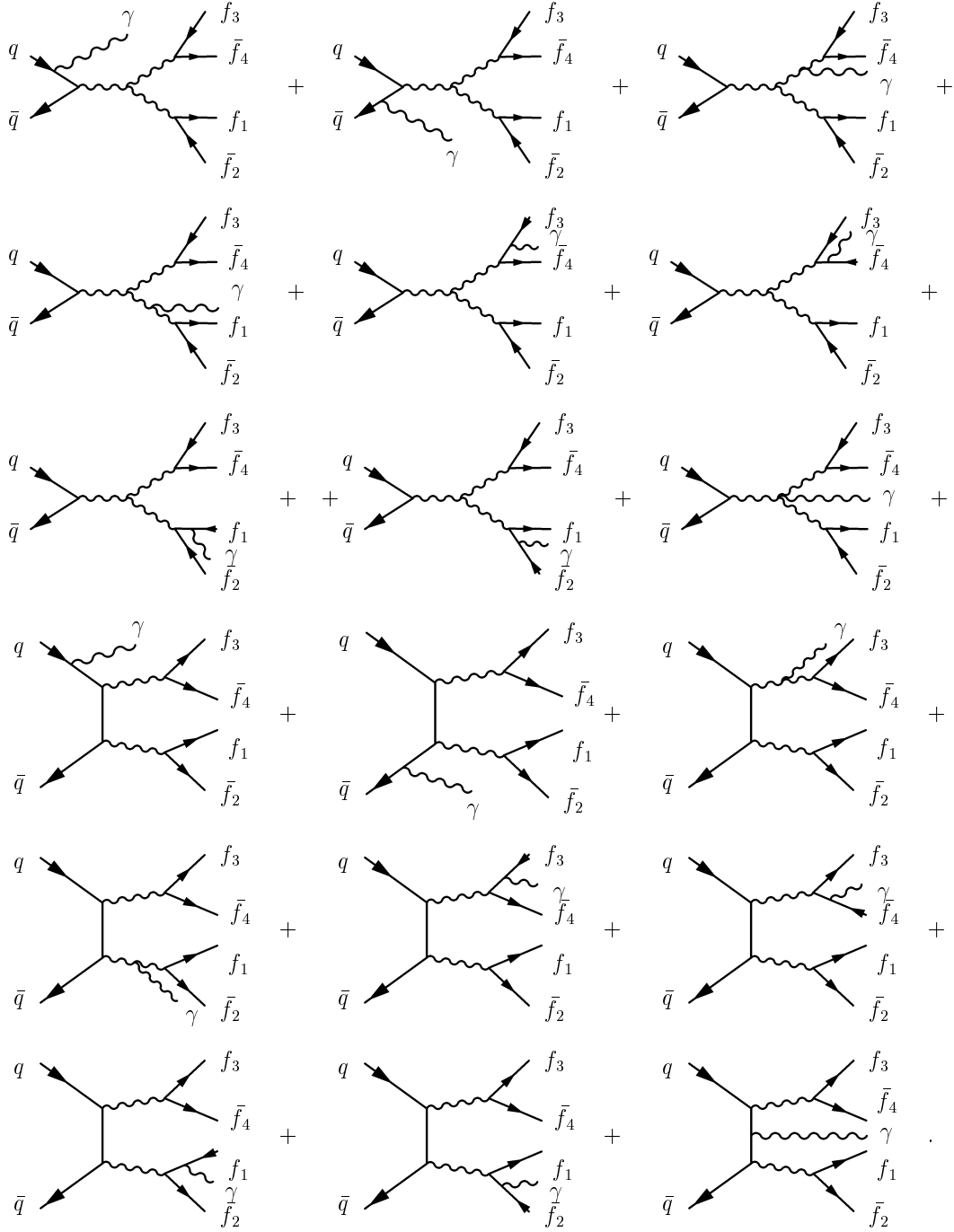}}
%\vskip -0.2cm
\caption{\label{all} Feynman diagrams for the process $ q \bar{q} \to W^+W^-\gamma \to f_1 \bar{f_2} f_3 \bar{f_4} 
\gamma$.}
\end{figure}

where for the s-channel each diagram stands for $\gamma$ as well as $Z$ exchange.

\subsection{The soft photon limit}

To get an analytical understanding of the underlying physics let's study first the soft photon limit, where 
the matrix element can be factorized

\be
\label{M}
{\cal M} = (-e) \; {\cal M}_0 \; \ep_{\kappa}^* (k) \;j^{\kappa}
\e

with ${\cal M}_0$ being the matrix element without photon radiation, the eikonal factor is

\bea
\label{Eikonal}
\fbox{\parbox{12.5cm} {
\bea
\quad j^{\kappa} &=& \quad \left(  Q_3 \, \frac{r_3^{\phantom{3}\kappa}}{r_3 \cdot k} \, + \, (1-Q_3) \, 
\frac{r_4^{\phantom{4}\kappa}}{r_4 \cdot k} \, - \, \frac{k_+^{\phantom{+}\kappa}}{k_+ \cdot k}\right)\, 
\frac{k_+^2-\overline{M}^2}
{(k_++k)^2-\overline{M}^2} \qquad \quad \qquad \no \\
 && - \left(  Q_1 \, \frac{r_1^{\phantom{1}\kappa}}{r_1 \cdot k} \, + \, (1-Q_1) \, 
\frac{r_2^{\phantom{2}\kappa}}{r_2 \cdot k} \, - \, \frac{k_-^{\phantom{-}\kappa}}{k_- \cdot k} \right)\, 
\frac{k_-^2-\overline{M}^2}
{(k_-+k)^2-\overline{M}^2} \no \\
 && + \left(- Q_q \, \frac{p_1^{\phantom{1}\kappa}}{p_1 \cdot k} \, - \, Q_{\bar{q}} \, \frac{p_2^{\phantom{2}\kappa}}{p_2 
\cdot k} \, + \, \frac{k_+^{\phantom{+}\kappa}}{k_+ \cdot k} \, - \, \frac{k_-^{\phantom{-}\kappa}}{k_- \cdot k} \right) \no 
\\
&& \no \\ 
 &=& \qquad D^{+\kappa} \, -\, D^{-\kappa} \, +\, P^{\kappa} \no \ea  }}  \qquad \quad
\ea

with $Q_i= |Q_i|$, $r_i$ being the momenta of the $f_i$ respectively and $ \overline{M } = M_{{\sub W}} - i \, \Gamma /2 $. 

Note also that we have made use of the partial fraction 

\bea
\label{partial} 
\frac{1}{k_{\pm}^2-\overline{M}^2}\, \frac{1}{(k_{\pm}+k)^2-\overline{M}^2} = \frac{1}{2\, k_{\pm} \cdot k} \, \left( \, 
\frac{1}{k_{\pm}^2-\overline{M}^2}\, - \, \frac{1}{(k_{\pm}+k)^2-\overline{M}^2} \right) \qquad
\ea

\vspace{0.5cm}
which can be illustrated as following:

%
%  Draw Diagram
%
\begin{figure}[H]
\vspace{-3cm}
\hspace{0.4cm}\centerline{\epsfysize=6cm\epsffile{parfrac.epsi}}
\end{figure}

\vspace{0.7cm}

To find a radiation zero one has to solve $ | \epsilon^*(k) \cdot j | ^2 \equiv 0$.

\subsection{The general case}

In the general case the cross section can be written as

\bea
\label{gen}
d \sigma &=& \frac{1}{2s} d \Phi_5 \,\left|\,\, \frac{{\cal M}_3+{\cal M}_5+{\cal M}_6+{\cal M}_{12}+{\cal M}_{14}+{\cal M}_{15}}{[(k_++k)^2-M_W^2-i\Gamma M_W] \,\,\, [k_-^2-M_W^2-i \Gamma M_W]}  \right.\vspace{0.4cm} \no \\  \no \\
&& \hspace{1.3cm} \, + \,  \frac{{\cal M}_4+{\cal M}_7+{\cal M}_8+{\cal M}_{13}+{\cal M}_{16}+{\cal M}_{17}}{[k_+^2-M_W^2-i\Gamma M_W] \,\,\,[ (k_-+k)^2-M_W^2-i \Gamma M_W]}\vspace{0.4cm} \no \\  \no \\
&& \hspace{0.8cm} \left. \, + \,  \frac{{\cal M}_1+{\cal M}_2+{\cal M}_3+{\cal M}_4+{\cal M}_9+{\cal M}_{10}+{\cal M}_{11}+{\cal M}_{12}+{\cal M}_{13}+{\cal M}_{18}}{[k_+^2-M_W^2-i\Gamma M_W]\,\,\, [k_-^2-M_W^2-i \Gamma M_W]}\,\, \right| ^2 \no \\
\ea

where the subscript refers to the diagrams of figure \ref{all}.
Integrating over the virtual momenta $k_{\pm}$ we find the five body phase space split into nine pieces which are all separated in phase space. Our actual interest is in the contribution where both $W's$ are on-shell, with $k_{\pm}^2=M_W^2$, 

\bea
d \Phi_5 = d \Phi_3 \, d \Phi_2^+ \, d \Phi_2^- \left( \frac{\pi}{\Gamma M_W} \right)^2
\ea

and only the last fraction of (\ref{gen}) will contribute, since no radiation off final fermions has to be considered.\\
 
Concerning the kinematics we choose to fix the direction of the $W^-$ by $\Theta$. The energy and the angle of the photon are given by
$E_\gamma$ and $\theta_\gamma$ respectively. Given the initial momenta $p_1$ and $p_2$ the $W^+$ momentum is constrained totally by 4-dim momentum conservation.
\bea
\label{kin2}
p_1^{\mu} & = & E(1,0,0,1) \no \\
p_2^{\mu} & = & E(1,0,0,-1) \no \\
k^{\mu} & = & E_{\gamma} (1, \sin\, \theta_{\gamma}, 0, \cos\, \theta_{\gamma}) \no \\
k_-^{\mu} & = & (E_W, \sqrt{E_W^2-M_W^2}\, \sin \, \Theta, 0, \sqrt{E_W^2-M_W^2}\, \cos \, \Theta) \no \\
k_+^{\mu} & = & p_1^{\mu}+p_2^{\mu}-k^{\mu}-k_-^{\mu}  \no \\
\ea

where $E_W$ is determined by the constraint $k_+^2=M_W^2$. Note that the scattering is restricted to the $x-z$ plane ($\phi_{\gamma}=0, \Phi=0$), which allows to study so-called type II zeros \cite{james}.

%Hence the differential cross section is given by

%\bea
%d \sigma &\equiv& \frac{d\, \sigma}{d\Omega^- \, d\Omega^+ \, d \cos \Theta \, d \Phi \, d \cos \theta_{\gamma} \, d 
%\phi_{\gamma} \, d \log E_{\gamma}} \no \\ \vspace{0.2cm}
%&= & 3 \, \frac{1}{(32 \pi^2)^2} \, \frac{1}{(2 \pi )^5} \, \frac{1}{2s} \, \frac{E_W}{2} \, 
%\frac{E_{\gamma}^2}{2} \, \frac{0.389 \,{\sub \times} \, 10^{12} \, \, \abs{\an{{\cal M}}}^2 
%}{\abs{-4E+2E_{\gamma}-2E_WE_{\gamma}/\cos(\theta_{\gamma}-\Theta)\sqrt{E_W^2-M_W^2}}^2}  \quad . \no \\ 
%\ea

%where we have denoted  $\an{{\cal M}}$ as the sum of ${\cal M}_i$'s as given in (\ref{crosssection}). \\

In the following plot we consider  the differential cross section as a function of $\cos \theta_{\gamma}$ and keep the other variables fixed using \cite{pdg}.\\

\begin{figure}[H]
\vspace{-1.2cm}
\centerline{\epsfysize=15cm\epsffile{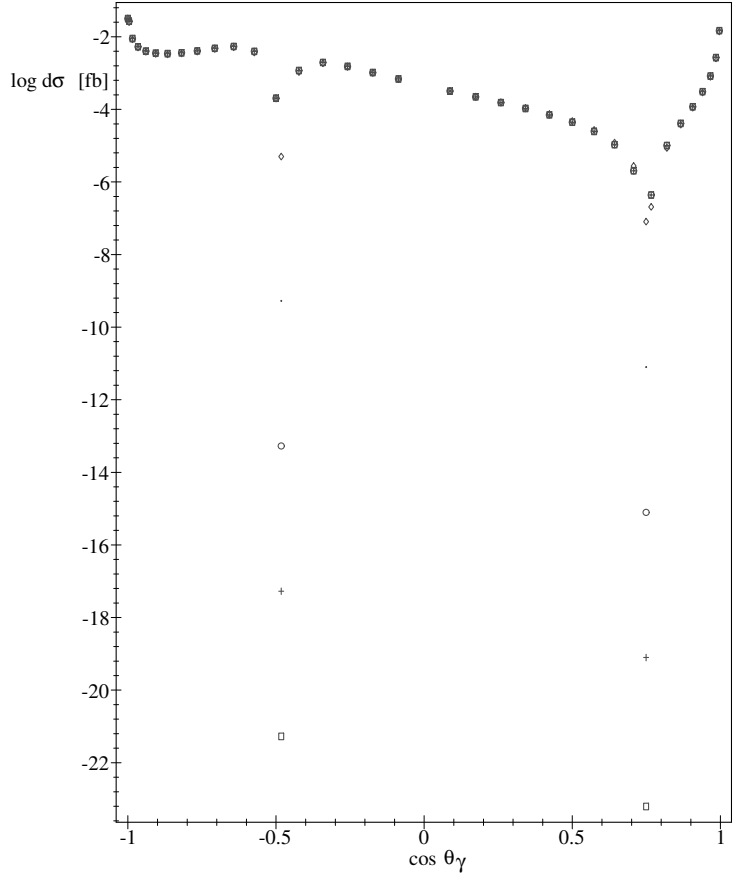}}
\vskip -1.2cm
\caption{\label{LRthetag}{Differential cross section $ d \sigma / d \Omega^+ d \Omega^- \; d \cos \!\, \Theta \; d \Phi \; d \cos \theta_{\gamma} \; d \phi_{\gamma} \; d \log  E_{\gamma}$ for the process $u \bar{u} \to W^+W^-\gamma \to f_1 \bar{f}_2 f_3 
\bar{f}_4 \gamma$ at $\sqrt{s}=1$ TeV with $\Theta=\frac{2\pi}{3}$, where $\diamond , \, {\sub +} , \, \circ , \, \cdot , \, \Box$ stand for $ E_{\gamma} = 10, \, 10^{-1}, \, 10^{-3}, 
10^{-5}, 10^{-7} $ GeV respectively.}}
\end{figure}

From figure \ref{LRthetag} we learn that only the softest photons will result in actual zeros, where higher photon energies only give rise to less and less deeper dips.

%\vspace{-0.5cm}
%\bea
%\label{para2}
%\begin{tabular}{lll}
%$ Q_q=\frac{2}{3} $, & $ E=500$ GeV, & $ \Theta = \frac{2\pi}{3} $. \\ & & 
%\end{tabular}
%\ea

\section{Anomalous gauge boson couplings}

In the previous sections we have considered standard model (SM) electroweak couplings and within this model we have been able to locate
various type II zeros for the process $q \bar{q} \to W^+W^-\gamma \to f_1\bar{f_2}f_3\bar{f_4}\gamma$. The question
of interest is whether those zeros will survive within the framework of non-standard couplings and if so whether their
positions will remain unchanged. \\

The trilinear and quartic couplings probe different aspects of the weak interactions. The trilinear couplings test the
non-Abelian gauge structure. In contrast, the quartic couplings can be regarded as a window on electroweak symmetry breaking.
The quartic couplings therefore present a connection to the scalar sector of the theory \cite{godfrey}. \\

It is quite possible that the quartic couplings deviate from their SM values while the triple gauge vertices do not.\\

If the mechanism for electroweak symmetry breaking does not reveal itself through the discovery of new particles such as the Higgs 
boson, supersymmetric particles or technipions it is quite possible that anomalous quartic couplings could be our first probes into 
this sector of the electroweak theory \cite{godfrey}. \\ 

We will therefore focus on the so-called genuine anomalous quartic couplings, i.e. those which give no contribution to the
trilinear vertices.\\

The lowest dimension operators which lead to genuine quartic couplings where at least one photon is involved are of dimension 6 \cite{belanger}. A
dimension 4 operator is not realised since a custodial $SU(2)$ symmetry is required to keep the $\rho$ parameter, $\rho = M_W^2/(M_Z^2 \cos ^2 \theta_w)$, to its measured SM value of $\rho=1$. Imposing SU(2) custodial symmetry the two Lagrangians giving rise to an 
anomalous $WW\gamma \gamma$ vertex are \cite{eboli}\\

\bea
{\cal L}_0 &=& - \frac{e^2}{16 \Lambda^2}\, a_0\, F^{\mu \nu} \, F_{\mu \nu} \vector{W^{\alpha}} \cdot \vector{W_{\alpha}} \no \\
{\cal L}_c &=& - \frac{e^2}{16 \Lambda^2}\, a_c\, F^{\mu \alpha} \, F_{\mu \beta} \vector{W^{\beta}} \cdot \vector{W_{\alpha}}  
\ea
with 
\bea
F^{\mu \nu}&=&  \partial _{\mu} A_{\nu} - \partial _{\nu} A_{\mu} \no \\
W_{\mu \nu}&=& \partial _{\mu} \vector{W}_{\nu} - \partial _{\nu} \vector{W}_{\mu} -g \vector{W}_{\mu} \times \vector{W}_{\nu}
\ea

and $\vector{W}_{\mu}$ being the vector

\bea
\label{wb}
\vector{W _{\mu}} = \left( \begin{array}{c}  \frac{1}{\sqrt{2}} ( W_{\mu}^+ + W_{\mu}^-) \\  \frac{i}{\sqrt{2}} ( W_{\mu}^+ - W_{\mu}^-) 
                                      \\  W_{\mu}^3 - \frac{g^{\prime}}{g} B_{\mu} \end{array} \right) = \left( \begin{array}{c}  \frac{1}{\sqrt{2}} ( W_{\mu}^+ + W_{\mu}^-) \\  \frac{i}{\sqrt{2}} ( W_{\mu}^+ - W_{\mu}^-) 
                                      \\  Z_{\mu}\, / \, \cos \theta_w \end{array} \right) \quad ,
\ea

The parameter $\Lambda$ is the scale at which new physics enters the game, as often in the literature we set its value to $M_W$. 
Any other choice of $\Lambda$ then simply corresponds to a rescaling of the couplings $a_0, a_c$ and $a_n$.
The anomalous contribution to the $WWZ\gamma $ vertex has its origin in the Largangian

\bea
{\cal L}_n &=&  - \frac{e^2}{16 \Lambda^2}\, a_n \epsilon_{ijk} W_{\mu \alpha}^{(i)} W_{\nu}^{(j)} W^{(k)\alpha} F^{\mu \nu} \quad .
\ea

In the following figure the differential cross section is plotted for various anomalous parameters $a_0, a_c$ and $a_n$. From the Lagrangian it can be seen that any anomalous contribution is linear in $E_\gamma$. Soft photons will only slightly be modified whereas for high energetic photons the anomalous contribution can be large. The task is to optimize the photon energy, to make it small enough to maintain the zero, but large enough to gain measurable deviations from the SM predictions.

\begin{figure}[H]
\vspace{-1.2cm}
\centerline{\epsfysize=15cm\epsffile{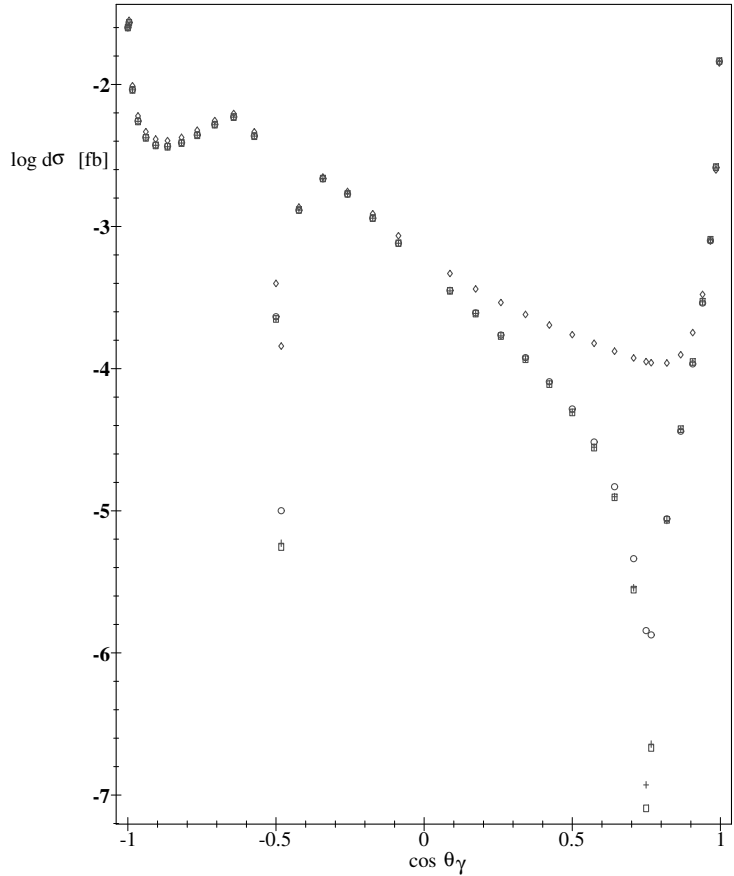}}
\vspace{-1.2cm}
\caption{{Differential cross section $ d \sigma / \, d \Omega^+ d \Omega^- \; d \cos \! \,\Theta \; d \Phi \;d \cos \theta_{\gamma} \; d \phi_{\gamma} \; d \log  E_{\gamma}$ for the process $ u \bar{u} \to W^+W^-\gamma \to f_1 \bar{f}_2 f_3 \bar{f}_4 \gamma $  at $\sqrt{s}=1$ TeV, with $\Theta=\frac{2\pi}{3}$ and $E_{\gamma}=1$ GeV where $\Box ,\,{\sub {+}}, \, \circ, \, \diamond $ stand for $ a_i= 0, \, 1,\, 10, \, 100 $.}}
\end{figure} 

%\begin{figure}[H]
%\vspace{-1.2cm}
%\centerline{\epsfysize=17cm\epsffile{AR1.eps}}
%\vspace{-1.2cm}
%\caption{{Differential cross section for the process $u \bar{u} \to W^+W^-\gamma \to f_1 \bar{f}_2 f_3 \bar{f}_4 \gamma$ with $E_{\gamma}%=1$ GeV where $\Box, \,{\sub +}, \, \circ, \, \diamond $ stand for $a_i= 0, \, 1,\, 10, \, 100$ for only right handed initial quarks.}}
%\end{figure} 

\section{Conclusion}
We have shown that the cross section is most sensitive to deviations from the standard model at the position of the radiation zeros, where we have in particular investigated anomalous quatric gauge boson couplings. The sensitivity to new physics of this type can even be increased operating with polarized beams, since the anomalous contribution originates in the four boson vertex, which only occurs in the $s$-channel, hence eliminating the $t$-channel using right handed quarks does have a considerable effect on the sensitvity.

\section*{Acknowledgements}
The work was fully supported by 'DAAD Doktorandenstipendium im Rahmen des gemeinsamen Hochschulsonderprogramms III von Bund und L\"andern'.

\end{document}